\documentclass[useAMS,usenatbib]{mn2e}
\usepackage{graphicx}
\usepackage{times}

\title[The multiplicity of Tr\,16-112, HD\,93343 and HD\,93250]{Optical spectroscopy of X-Mega targets in the Carina nebula - VII. On the multiplicity of Tr\,16-112, HD\,93343 and HD\,93250\thanks{Based on observations collected at the European Southern Observatory (La Silla, Chile), at Complejo Astron\'omico El Leoncito (Argentina), at the Cerro Tololo Inter-American Observatory (CTIO) and with {\it XMM-Newton}, an ESA Science Mission with instruments and contributions directly funded by ESA Member States and the USA (NASA).}}

\author[G.\ Rauw et al.]{G.\ Rauw$^1$\thanks{Research Associate FRS/FNRS (Belgium)}, Y.\ Naz\'e$^1$\thanks{Postdoctoral Researcher FRS/FNRS (Belgium)}, E.\ Fern\'andez Laj\'us$^2$, A.A.\ Lanotte$^1$, G.R.\ Solivella$^2$,\newauthor 
H.\ Sana$^3$, and E.\ Gosset$^1$\thanks{Senior Research Associate FRS/FNRS (Belgium)}\\
$^1$ Institut d'Astrophysique \& G\'eophysique, Universit\'e de Li\`ege, B\^at.\ B5c, All\'ee du 6 Ao\^ut 17, B-4000 Li\`ege, Belgium\\
$^2$ Facultad de Ciencias Astron\'{o}micas y Geof\'{\i}sicas, Universidad Nacional de La Plata, Paseo del Bosque S/N, 1900 La Plata, Argentina\\
$^3$ European Southern Observatory, Alonso de Cordova 3107, Vitacura, Santiago 19, Chile}

\date{Accepted date
      Received date;
      in original form \today}
\pagerange{\pageref{firstpage}--\pageref{lastpage}}
\pubyear{2009}

\begin{document}

\maketitle

\label{firstpage}

\begin{abstract}
We present the results of a spectroscopic monitoring campaign devoted to three O-type stars in the Carina nebula. We derive the full SB2 orbital solution of the binary system Tr\,16-112, an exceptional dissymmetrical system consisting of an O5.5-6\,V((f$^+$?p)) primary and a B2\,V-III secondary. We also report on low-amplitude brightness variations in Tr\,16-112 that are likely due to the ellipsoidal shape of the O5.5-6 primary revolving in an eccentric orbit around the system's centre of mass. We detect for the first time a clear SB2 binary signature in the spectrum of HD\,93343 (O8 + O8), although our data are not sufficient to establish an orbital solution. This system also displays low amplitude photometric modulations. On the other hand, no indication of multiplicity is found in the optical spectra of HD\,93250. Finally, we discuss the general properties of multiple massive stars in the Carina OB1 association.
\end{abstract}


\section{Introduction}
The investigation of the multiplicity of early-type stars in very young open clusters is a powerful tool to better understand many problems related to massive star research. First of all, such studies allow us to enlarge the - still rather small - sample of massive stars with well determined fundamental properties such as masses and radii. Moreover, the statistical properties of the binary systems also constrain the formation scenarios of massive stars (see e.g. \citealt{Sana08}).

The region around $\eta$ Carinae harbours several very young open clusters and is therefore extremely rich in early-type stars. As a consequence, this area has been the subject of many investigations over the last thirty years. In this context and in the framework of the so-called {\it X-Mega} project \citep{Corcoran}, aiming (among other things) at an intensive monitoring of the region around $\eta$ Carinae with the {\it ROSAT} X-ray satellite, a large multi-wavelength observing effort was devoted to the Carina Nebula and the Trumpler 16 cluster in particular. One of the prime objectives was to obtain accurate ephemerides of colliding-wind early-type binary systems to interpret their X-ray light curves. The spectroscopic monitoring of these objects started in the late 1990's and led to a series of papers \citep{ALB01,ALB02,Morrell,GR01,Naze,Niemela} reporting the results of dedicated investigations of the properties of a number of O-type binaries in the Carina Nebula. This paper is the seventh in this series and is devoted to three objects: the binary systems Tr\,16-112 and HD\,93343 as well as the likely single star HD\,93250. Furthermore, we briefly discuss the general properties of O-type binaries in and around Trumpler\,16 as derived in the framework of the present campaign and other related studies. 

\section{Observations}
\subsection{Medium resolution spectroscopy}
Six medium resolution spectra of Tr\,16-112 and of HD\,93343 in the wavelength range $3850 - 4800$\,\AA\ were gathered in 1997 with the ESO 1.5\,m telescope equipped with a Boller \& Chivens (B\&C) Cassegrain spectrograph (see \citealt{GR01} for a detailed description of the instrumentation and the data reduction). 

Several spectra of Tr\,16-112 were obtained in 1986 with the Cassegrain spectrograph at the CTIO 1\,m telescope equipped with a two-dimensional, photon-counting detector (2D-frutti). The wavelength coverage was from 3800 to 5000\,\AA\ with a 3 pixel resolution of 1.5\,\AA. The S/N ratios were about 50.
Several low-resolution spectra of the same star were obtained in 1993, 2007 and 2008 at Complejo Astron\'omico El Leoncito (Casleo)\footnote{Casleo is operated under agreement between CONICET and the National Universities of La Plata, C\'ordoba and San Juan.}, Argentina. The spectra were taken with the modified REOSC SEL\footnote{Spectrograph Echelle Li\`ege (jointly built by REOSC and Li\`ege Observatory and on long term loan from the latter).} spectrograph in simple dispersion Cassegrain mode attached to the 2.15\,m Jorge Sahade telescope. The dispersion was 1.64\,\AA\ per pixel. The data cover the wavelength range from 3800 to 4800\,\AA\ and have a S/N of about 200. The spectra were reduced at Facultad de Ciencias Astron\'omicas y Geof\'{\i}sicas (Universidad Nacional de La Plata) with the {\sc iraf}\footnote{{\sc iraf} is distributed by NOAO, operated by AURA, Inc., under agreement with NSF.} software version 2.10. Radial velocities of the He\,{\sc ii} $\lambda$\,4542 and $\lambda$\,4686 lines were obtained by fitting a Gaussian profile to the 
observed lines.
\subsection{High resolution echelle spectroscopy}
During several observing campaigns (May 1999, May 2000, May 2001, March \& April 2002 and May 2004), we took a series of echelle spectra of our targets (27 for Tr\,16-112, 9 for HD\,93343 and 6 for HD\,93250) with the Fiber-fed Extended Range Optical Spectrograph (FEROS) mounted first on the ESO 1.5\,m (until October 2002) and later on the ESO/MPG 2.2\,m telescopes at La Silla. These spectra cover the wavelength domain from about 3750 to 9000\,\AA\ with a resolving power of 48000. The detector was an EEV CCD with $2048 \times 4096$ pixels of size $15 \times 15$\,$\mu$m$^2$. The exposure times were typically 30\,min and the typical S/N ratios were about 150. The data were reduced using the specific context under the {\sc midas} environment along with an improved reduction pipeline (see \citealt{Sana03}). Finally, the FEROS spectra were normalized to the continuum using polynomials.

High-resolution echelle spectra were also obtained at Casleo in 1994 and 2008 with the REOSC SEL instrument in echelle mode attached to the 2.15\,m telescope. The detector was a Tek $1024 \times 1024$ pixel CCD. The spectra cover the wavelength range from 3600 to 6000\,\AA\ with a spectral resolving power of 18000. Typical S/N ratios are about 80. The REOSC echelle spectra were reduced using {\sc iraf} routines. Individual echelle orders were normalized using carefully chosen continuum windows.\\

\subsection{Photometry of Tr\,16-112}
CCD photometry of Tr\,16-112 was performed during the nights of February 21-27, 2007, using the CH-250 camera with a PM512 CCD chip, attached to the 0.6\,m Helen Sawyer Hogg telescope (f/15 Cassegrain) at Casleo. We used a standard Johnson-Cousins $B\,V\!R\,I$ filter set for the observations, although the bulk of the data were taken in the $V$ band. The CCD images were calibrated in the  standard way using bias, dark and flat-field frames and finally the photometry was derived using the APPHOT/IRAF package. The typical uncertainty on the differential $V$-band photometry of Tr\,16-112 is about 0.005\,mag. 

\section{Interstellar lines}
We added up all the FEROS spectra of our targets in the wavelength domain 3870 -- 3950\,\AA\ to derive high quality spectra of the interstellar He\,{\sc i} $\lambda$\,3889 and Ca\,{\sc ii} K lines. For all three stars, the profiles of the Ca\,{\sc ii} K lines look remarkably similar to the plots given by \citet{Wal82}. The high signal to noise ratio of our mean spectra allows us to further detect some weak (equivalent widths of a few m\AA) higher velocity components that were not seen in the older data. In the case of Tr\,16-112, we definitely find a component at $-143$\,km\,s$^{-1}$ and possibly another one at $+205$\,km\,s$^{-1}$. In HD\,93343, we clearly identify an absorption at $-75$\,km\,s$^{-1}$, whilst in HD\,93250 we detect a component at $-98$\,km\,s$^{-1}$. All three stars display a moderately prominent interstellar He\,{\sc i} $\lambda$\,3889 absorption line at a heliocentric velocity of $-28.0$\,km\,s$^{-1}$ (for Tr\,16-112 and HD\,93250) and $-25.4$\,km\,s$^{-1}$ (HD\,93343). The equivalent widths of these features are $0.010$ (HD\,93343), $0.045$ (HD\,93250) and $0.080$\,\AA\ (Tr\,16-112).

\section{The O + B binary Tr\,16-112}
Tr\,16-112 ($\equiv$ CPD\,$-59\degr$\,2641 $\equiv$ CD\,$-59\degr$\,3310, \citealt*{FMM}) is a member of the Tr\,16 cluster. Based on 10 low-resolution spectra, \citet{Levato} classified Tr\,16-112 as an SB1 binary system with an orbital period of 4.02\,days. Their preliminary orbital solution yielded a significant eccentricity of 0.23. \citet{Luna} subsequently obtained 5 additional RV data points from echelle spectra and re-computed an SB1 orbital solution with a period of 4.080\,days and $e = 0.25 \pm 0.03$. 

\subsection{The spectral types}
We have applied a disentangling method based on the principles discussed by \citet{GL06} to the normalized FEROS spectra of Tr\,16-112. For this purpose, we have selected two wavelength domains within the spectral range covered by the FEROS data: our violet and blue domains cover respectively the wavelength intervals [4000, 4250] and [4410,4790]. Disentangling the spectra of the components of this star was a challenging task that pushed our code to its limits. Indeed, the secondary star contributes only about 8 percent of the total light of the system making it rather difficult to accurately follow its spectral signature. 

\begin{figure*}
\begin{minipage}{8.5cm}
\resizebox{8.5cm}{!}{\includegraphics{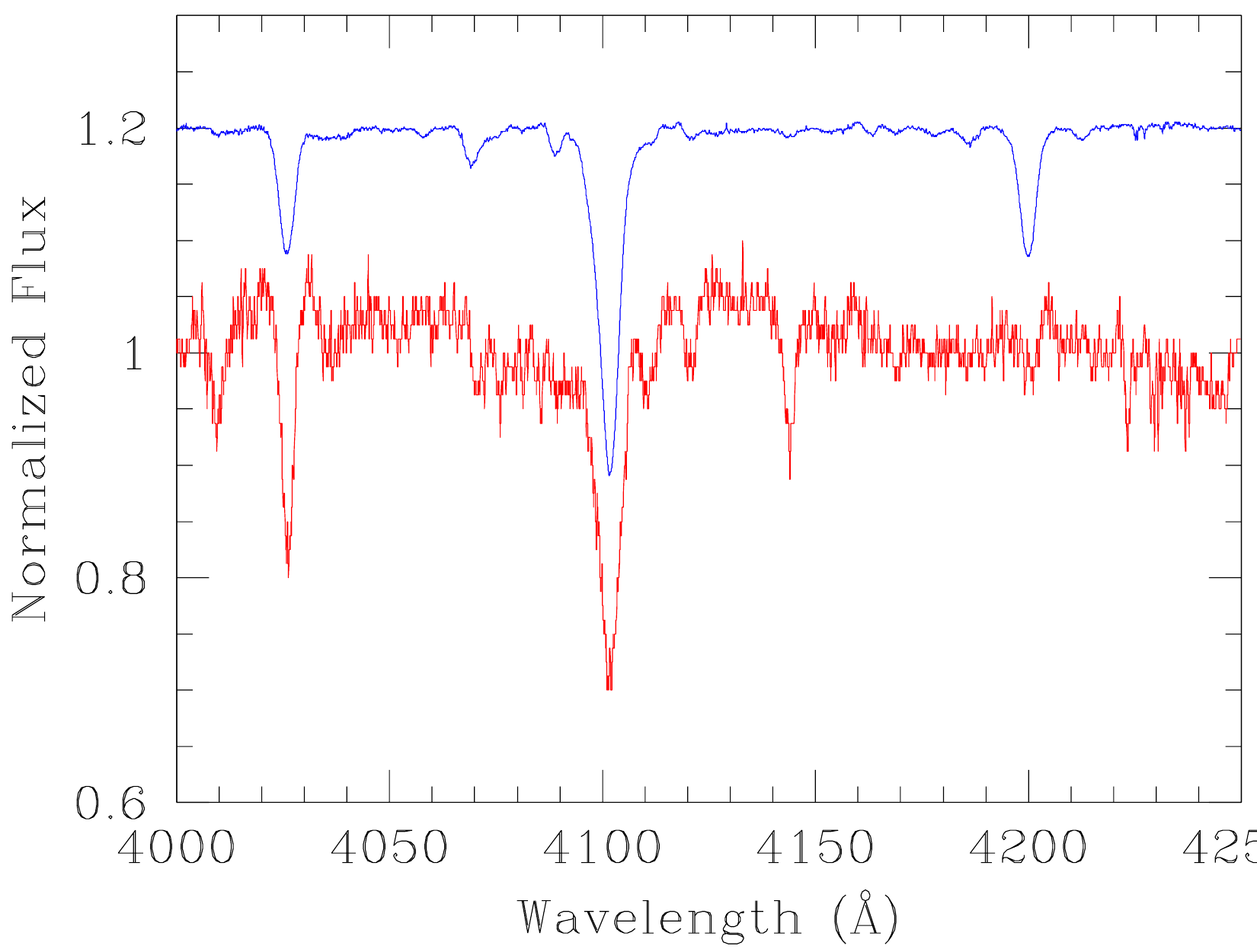}}
\end{minipage}
\hfill
\begin{minipage}{8.5cm}
\resizebox{8.5cm}{!}{\includegraphics{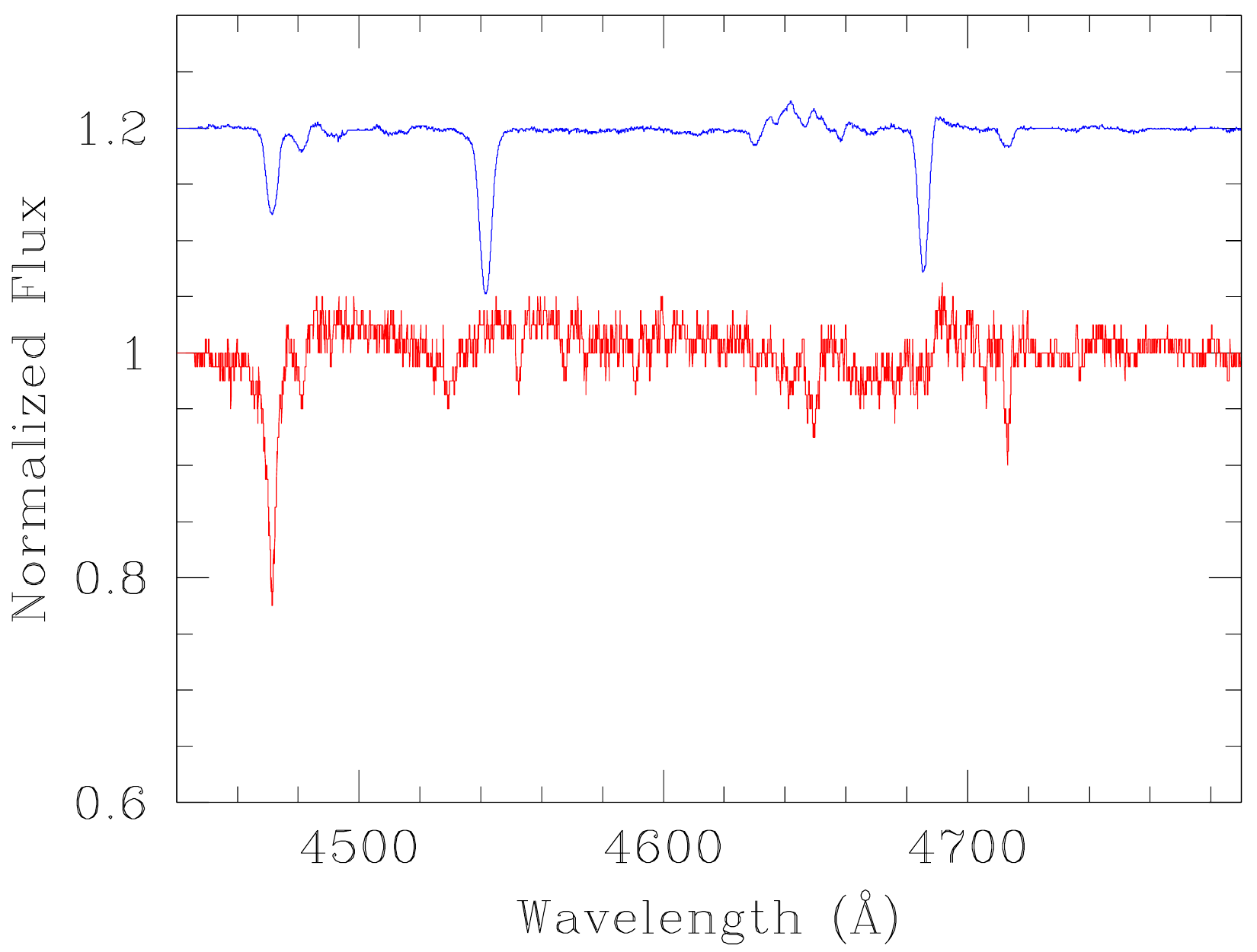}}
\end{minipage}
\caption{The normalized disentangled violet (left) and blue (right) spectra of the primary and secondary components of Tr\,16-112. The disentangled spectra were normalized accounting for a brightness ratio (secondary/primary) of 0.09. For clarity, the primary spectrum was shifted upwards by 0.2 continuum units.\label{disentspec}}
\end{figure*}

The resulting separated spectra are displayed in Fig.\,\ref{disentspec}. Although the disentangled secondary spectrum is admittedly of poor quality, these spectra reveal some interesting features. The apparent equivalent widths (i.e.\ with respect to the combined primary + secondary continuum) of the He\,{\sc i} $\lambda$\,4471 line as determined from the mean disentangled spectra are $0.325 \pm 0.005$ and 0.073\,\AA\ for the primary and secondary respectively. For the He\,{\sc ii} $\lambda$\,4542 and $\lambda$\,4686 lines, the apparent primary's equivalent widths (EWs) are $0.668 \pm 0.005$ and 0.426\,\AA\ respectively. None of the latter lines is seen in the secondary spectrum (see Fig.\,\ref{disentspec}). 

The He\,{\sc i} $\lambda$\,4471/He\,{\sc ii} $\lambda$\,4542 EW ratio ($\log{W'} = -0.313 \pm 0.007$, see \citealt{Conti73b} and \citealt{Mathys}) places the primary star at the border between spectral types O5.5 and O6. The disentangled primary spectrum reveals a weak, but definite, emission blend of N\,{\sc iii} $\lambda\lambda$\,4634-41 as well as a comparably strong C\,{\sc iii} $\lambda\lambda$\,4647-50 emission and a weak C\,{\sc iv} $\lambda$\,4658 P-Cygni profile\footnote{From its wavelength, we can actually rule out that the absorption component of this P-Cygni profile would be due to a blend with the Si\,{\sc iv} $\lambda$\,4654 absorption.}. The fact that the strength of the C\,{\sc iii} $\lambda\lambda$\,4647-50 and N\,{\sc iii} $\lambda\lambda$\,4634-41 blends are very much comparable would put the primary into the scarce category of Of?p stars (see \citealt{Wal72,Wal73} and \citealt*{Naze2}). We caution however that the known members of the Of?p category exhibit supergiant-like features and often display strong spectroscopic variability (see \citealt{Naze2}). In our data, we find no evidence for such features in the spectrum of the primary star of Tr\,16-112. The He\,{\sc ii} $\lambda$\,4686 absorption displays a slightly asymmetric profile with some hints of a weak P-Cygni type emission component. Some evidence for weak S\,{\sc iv} $\lambda\lambda$\,4486, 4504 emissions is found as well as a very weak Si\,{\sc iv} $\lambda$\,4116 emission, whilst Si\,{\sc iv} $\lambda$\,4089 is clearly seen in absorption. These features motivate a spectral classification of the primary as O5.5-6\,V((f$^+$?p)). 

Using the Fourier transform method (see \citealt{SimonDiaz}), we have determined the projected rotational velocity $v\,\sin{i} = (170 \pm 5)$\,km\,s$^{-1}$ from the He\,{\sc ii} $\lambda\lambda$\,4200, 4542 and O\,{\sc iii} $\lambda$\,5592 lines in the disentangled primary spectrum. This value is slightly lower than the values previously determined from the composite {\it IUE} spectra of the system ($195$ and $184$\,km\,s$^{-1}$, see \citealt{Penny} and \citealt{Howarth} respectively). One reason for this difference could be blending effects with the secondary's spectrum in the {\it IUE} data, as Penny mentioned the possible presence of the secondary signature in the cross-correlation function.    

The lack of He\,{\sc ii} lines in the spectrum of the secondary indicates that this is a B-type star. The most prominent features in the disentangled secondary spectrum are the He\,{\sc i} lines ($\lambda\lambda$\,4009, 4026, 4121, 4144, 4471, 4713) as well as H$\delta$. Somewhat weaker features are due to the C\,{\sc iii} + O\,{\sc ii} blends at 4070 and 4650\,\AA, as well as the Si\,{\sc iii} $\lambda\lambda$\,4552, 4568, Mg\,{\sc ii} $\lambda$\,4481, N\,{\sc ii} $\lambda$\,4530 and O\,{\sc ii} $\lambda$\,4110 lines. The presence of a very weak Si\,{\sc iv} $\lambda$\,4089 absorption cannot be ruled out. The relative strength of the lines of the silicon ions and the strength of the Mg\,{\sc ii} line compared to He\,{\sc i} $\lambda$\,4471 suggest a spectral type B2 (or slightly earlier) with an uncertainty of about half a subclass (see \citealt{WF}). The relative strength of the N\,{\sc ii} $\lambda$\,4530 line argues for a V-III luminosity class (see the luminosity effects at spectral type B2 in \citealt{WF}).

To evaluate the brightness ratio between the components of the system, we have compared the equivalent widths of prominent lines as measured in the combined spectrum to the typical EWs of presumably single stars of the same spectral type. For the primary, we used the He\,{\sc i} $\lambda\lambda$\,4026, 4471 as well as He\,{\sc ii} $\lambda\lambda$\,4200, 4542 lines and we took the typical values of the EWs from the paper of \citet{Conti73a}. For the secondary, our comparison was based on the He\,{\sc i} $\lambda\lambda$\,4026, 4471 and Mg\,{\sc ii} $\lambda$\,4481 lines and the typical values of the EWs of B2\,V stars given by \citet{Didelon}. This comparison shows that the secondary contributes about $8 \pm 3$ percent of the total light of Tr\,16-112 in the blue-violet spectral domain (i.e.\ the brightness ratio secondary/primary amounts to 0.09). 

The disentangling code offers the possibility of simultaneously determining the radial velocities (RVs) of the binary components. For this purpose, we used the lines listed in Table\,\ref{mask}. 
\begin{table}
\begin{center}
\caption{Spectral lines used for the radial velocity determination within the disentangling procedure.\label{mask}}
\begin{tabular}{c c}
Primary & Secondary \\
\hline
\multicolumn{2}{c}{Violet domain} \\
\hline
He\,{\sc i} $\lambda$\,4026 & He\,{\sc i} $\lambda$\,4026 \\
Si\,{\sc iv} $\lambda$\,4089 &  \\
H\,{\sc i} $\lambda$\,4102 & H\,{\sc i} $\lambda$\,4102 \\
He\,{\sc ii} $\lambda$\,4200 & \\
\hline
  \multicolumn{2}{c}{Blue domain} \\
\hline
He\,{\sc i} $\lambda$\,4471 & He\,{\sc i} $\lambda$\,4471 \\
Mg\,{\sc ii} $\lambda$\,4481 & Mg\,{\sc ii} $\lambda$\,4481 \\
He\,{\sc ii} $\lambda$\,4542 & \\
& Si\,{\sc iii} $\lambda$\,4553 \\
& Si\,{\sc iii} $\lambda$\,4568 \\
& O\,{\sc ii} $\lambda$\,4591 \\
He\,{\sc ii} $\lambda$\,4686 & \\
He\,{\sc i} $\lambda$\,4713 & He\,{\sc i} $\lambda$\,4713 \\
\hline
\end{tabular}
\end{center}
\end{table}

\subsection{The orbital solution}
Since the spectral signature of the secondary component of Tr\,16-112 is rather weak, we started by treating the system as an SB1 binary. We thus measured the radial velocities (RVs) of the primary component by fitting gaussian line profiles to the strongest lines. It was only on a few FEROS spectra that we directly noted the presence of the secondary's spectral signature in the He\,{\sc i} lines. In the latter cases, we simultaneously fitted two gaussian profiles. Since the secondary is of a significantly later spectral type than the primary, the He\,{\sc ii} lines were not affected and we thus consider that the radial velocities of the He\,{\sc ii} lines best reflect the motion of the primary of Tr\,16-112. In this way, we obtained a time series of 48 RV measurements from high-resolution echelle spectra, spanning almost 9 years (3259 days). Note that the RV dispersion of the diffuse interstellar band (DIB) at 5780\,\AA\ in the FEROS spectra of Tr\,16-112, was found to be 2.3\,km\,s$^{-1}$. This value can be seen as an estimate of the uncertainty on the RV of a constant spectral line in the spectrum of this star.

\begin{table}
\caption{Radial velocities of Tr\,16-112 as measured on our high-resolution echelle spectra. The first column yields the date of the observation. The second column indicates the primary RV as inferred from the direct measurement of the He\,{\sc ii} lines. Columns 3 and 4 yield the primary and secondary RVs as inferred from the disentangling method. Finally, the last column indicates the instrument that was used for the observation: FEROS (F) or the Casleo (C) SEL echelle spectrograph.\label{RVs}}
\begin{tabular}{c r r r c}
\hline
HJD$-$2450000 & RV$_p$ (He\,{\sc ii}) & RV$_p$ & RV$_s$ & Inst.\\
            & (km\,s$^{-1}$) & (km\,s$^{-1}$) & (km\,s$^{-1}$) & \\
\hline
 1299.624  &   95.9  &    90.3  &   $-374.8$ & F \\
 1300.630  &    7.1  &     2.2  &     $-7.7$ & F \\
 1301.562  & $-79.2$ &  $-82.3$ &     214.9  & F \\
 1301.724  & $-77.2$ &          &            & F \\
 1302.583  & $-44.5$ &  $-49.9$ &     123.4  & F \\
 1304.606  &    3.1  &     1.5  &    $-17.3$ & F \\
 1327.643  &  103.5  &   101.8  &   $-317.9$ & F \\
 1669.632  &   62.9  &    65.9  &   $-217.0$ & F \\
 1670.579  & $-57.5$ &          &            & F \\
 1672.568  &   45.5  &    46.2  &   $-171.9$ & F \\
 2037.575  & $-22.2$ &  $-27.9$ &     101.1  & F \\
 2037.690  &  $-2.7$ &          &            & F \\ 
 2038.531  &  107.1  &   101.7  &   $-370.5$ & F \\
 2039.594  &  $-5.3$ &          &            & F \\  
 2040.603  & $-72.7$ &  $-79.7$ &     287.4  & F \\
 2040.655  & $-74.7$ &  $-78.9$ &     268.3  & F \\
 2335.682  &  105.4  &    96.9  &   $-382.6$ & F \\
 2337.661  & $-78.4$ &  $-82.3$ &     260.2  & F \\
 2338.672  & $-33.1$ &  $-39.7$ &     105.8  & F \\
 2339.674  &  102.2  &    96.5  &   $-384.8$ & F \\
 2382.533  & $-58.9$ &  $-63.4$ &     208.3  & F \\
 2383.536  &   73.7  &    72.1  &   $-257.6$ & F \\
 3130.530  &   77.7  &    72.5  &   $-284.1$ & F \\
 3131.531  &   37.8  &    40.0  &   $-156.1$ & F \\
 3132.524  & $-69.0$ &  $-70.7$ &     234.9  & F \\
 3133.512  & $-60.5$ &          &            & F \\  
 3134.540  &   78.6  &    75.0  &   $-276.7$ & F \\
 3135.514  &   38.9  &    36.1  &   $-177.3$ & F \\
 4474.780 & $-59.4$ &          &            & C \\ 
 4474.802 & $-61.6$ &          &            & C \\ 
 4475.704 &   67.6  &          &            & C \\
 4475.726 &   63.8  &          &            & C \\
 4476.723 &   74.8  &          &            & C \\
 4476.745 &   55.0  &          &            & C \\
 4477.725 & $-62.5$ &          &            & C \\
 4477.747 & $-92.2$ &          &            & C \\
 4524.780 &   66.3  &          &            & C \\
 4524.803 &   64.7  &          &            & C \\
 4526.513 & $-66.9$ &          &            & C \\
 4526.539 & $-78.7$ &          &            & C \\
 4555.567 &  $-4.3$ &          &            & C \\
 4555.589 &    7.4  &          &            & C \\
 4556.556 &  113.8  &          &            & C \\
 4556.578 &  104.0  &          &            & C \\
 4557.598 & $-36.7$ &          &            & C \\
 4557.620 & $-33.8$ &          &            & C \\
 4558.626 & $-80.5$ &          &            & C \\
 4558.648 & $-79.2$ &          &            & C \\
\hline
\end{tabular}
\end{table}

\begin{figure}
\resizebox{8.5cm}{!}{\includegraphics{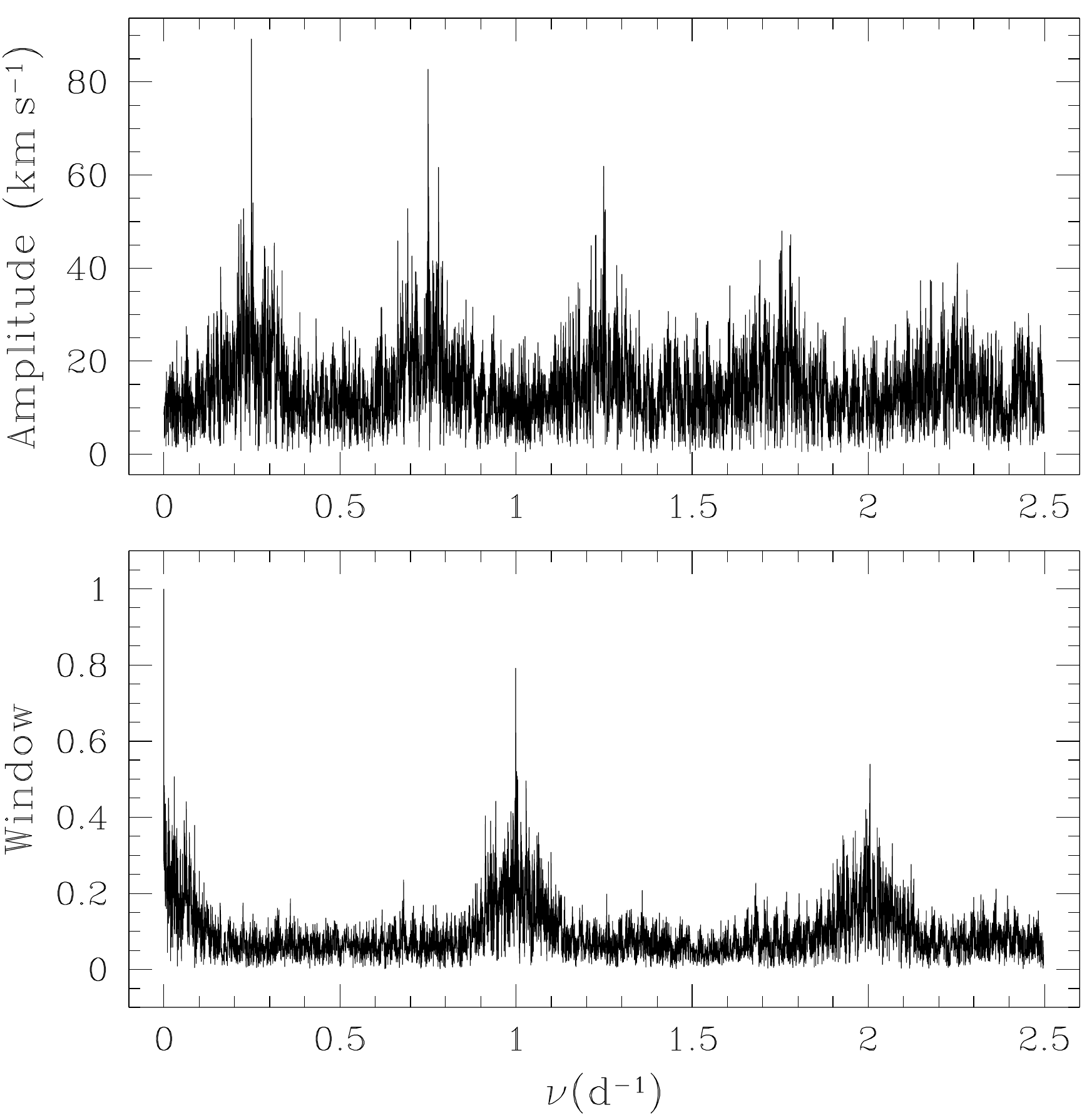}}
\caption{Top panel: periodogram (semi-amplitude spectrum) of the entire time series of 98 radial velocities of the primary component of Tr\,16-112 computed with the method of \citet{HMM}. Bottom panel: spectral window corresponding to our time series.\label{Fourier}}
\end{figure}
We then used the generalized periodogram Fourier technique discussed by \citet*{HMM} and \citet{Gosset} as well as the trial period method of \cite{LK} to perform a period search on the primary radial velocities. Previous attempts to determine the orbital period from a subset of the data discussed here revealed a severe ambiguity between a period of almost exactly 4.0 days and its alias at 1.3 days. To get rid of the aliasing problem, we thus obtained several spectra during the same night in the course of the May 2001 FEROS campaign and during the Casleo campaigns in 2008. The periodogram built from the new FEROS + Casleo data now clearly reveals its highest peak at a frequency of $(0.24902 \pm 0.00003)$\,d$^{-1}$ corresponding to an orbital period of $(4.0157 \pm 0.0005)$\,days. Our value is in excellent agreement with the 4.02\,day period originally proposed by \citet{Levato}. 

To check this result, we added a set of 56 older, lower resolution, RV measurements to our high-resolution data, thereby extending the time span of our data set to 24 years (8783 days). The analyses of this extended time series clearly confirm the conclusions from our high-resolution data set (see Fig.\,\ref{Fourier}), slightly reducing the uncertainty on the period to $(4.0157 \pm 0.0002)$\,days.   

\begin{table*}
\caption{Orbital parameters of the SB1 (left) and SB2 (right) solutions of Tr\,16-112. $T_0$ yields the time of periastron passage. The quoted uncertainties correspond to the $1\,\sigma$ error bars. \label{solorb}}
\begin{tabular}{l c c c c}
\hline
& SB1 & & \multicolumn{2}{c}{SB2} \\ 
\cline{2-2}\cline{4-5}
                       & Primary        & & Primary        & Secondary \\
Period (days)          & 4.0157 (fixed) & & \multicolumn{2}{c}{4.0157 (fixed)}\\
$\gamma$ (km\,s$^{-1}$)& $-1.3 \pm 1.0$ & & $-4.9 \pm 1.5$ & $-2.9 \pm 2.9$ \\
$K$ (km\,s$^{-1}$)     & $94.3 \pm 1.5$ & & $91.7 \pm 1.5$ & $319.7 \pm 5.1$ \\ 
$e$                    & $0.16 \pm 0.02$ & & \multicolumn{2}{c}{$0.15 \pm 0.01$}\\ 
$\omega$ ($^\circ$)    & $1.1 \pm 5.6$   & & \multicolumn{2}{c}{$342.8 \pm 5.6$}\\ 
$a\,\sin{i}$ (R$_{\odot}$) & $7.38 \pm 0.12$ & & $7.19 \pm 0.12$ & $25.05 \pm 0.40$ \\
$T_0$ (HJD) & $2454560.504 \pm 0.059$ & & \multicolumn{2}{c}{$2454560.310 \pm 0.060$}\\
$f(m)$ (M$_{\odot}$) & $0.335 \pm 0.016$ & & & \\
$m\,\sin^3{i}$ (M$_{\odot}$) & & & $21.7 \pm 0.9$ & $6.2 \pm 0.2$\\
$\overline{|O - C|}$ (km\,s$^{-1}$) & 4.7 & & 3.8 & 12.4 \\
\hline
\end{tabular}
\end{table*}  
\begin{figure*}
\begin{minipage}{8.5cm}
\resizebox{8.5cm}{!}{\includegraphics{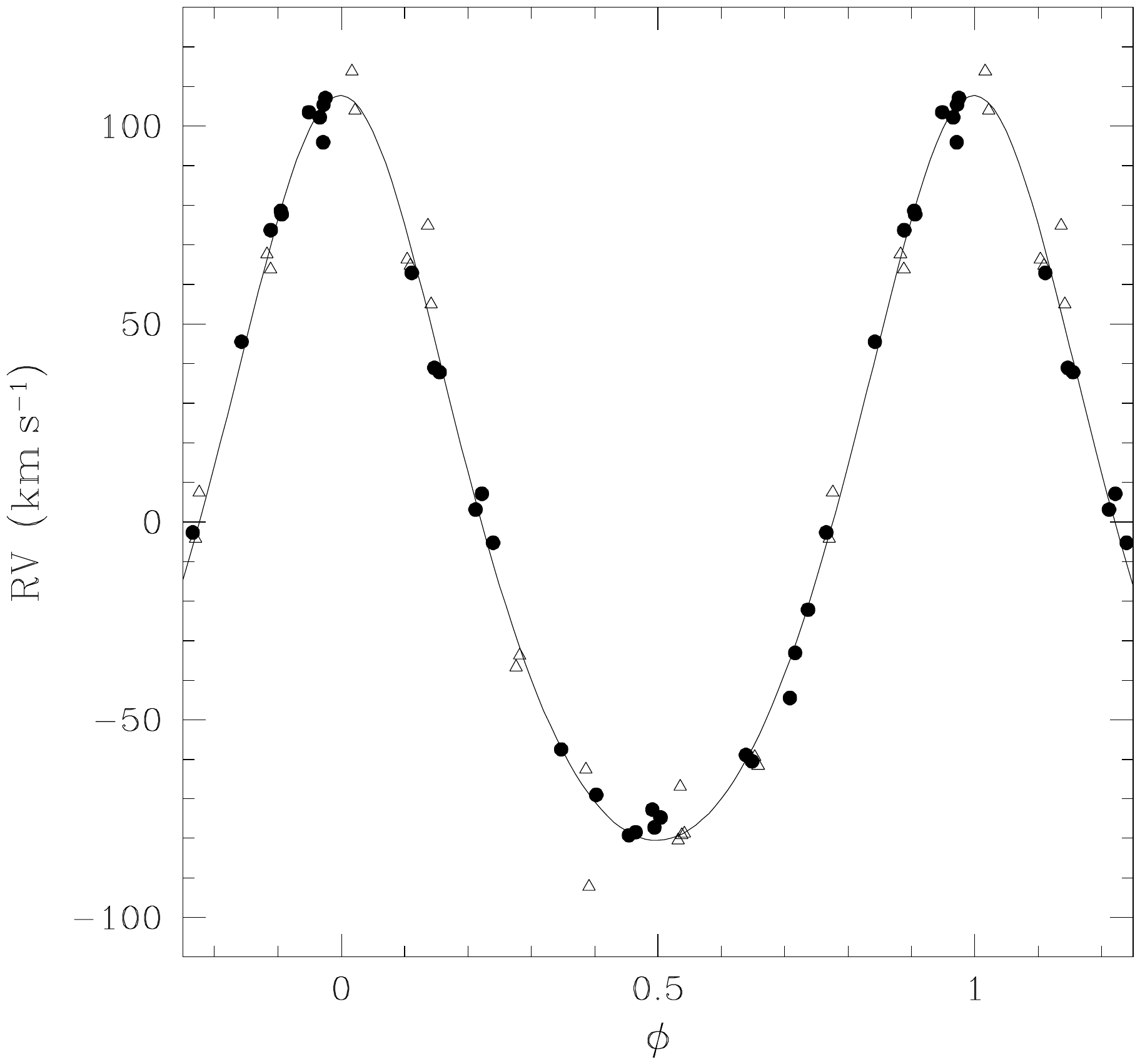}}
\end{minipage}
\hfill
\begin{minipage}{8.5cm}
\resizebox{8.5cm}{!}{\includegraphics{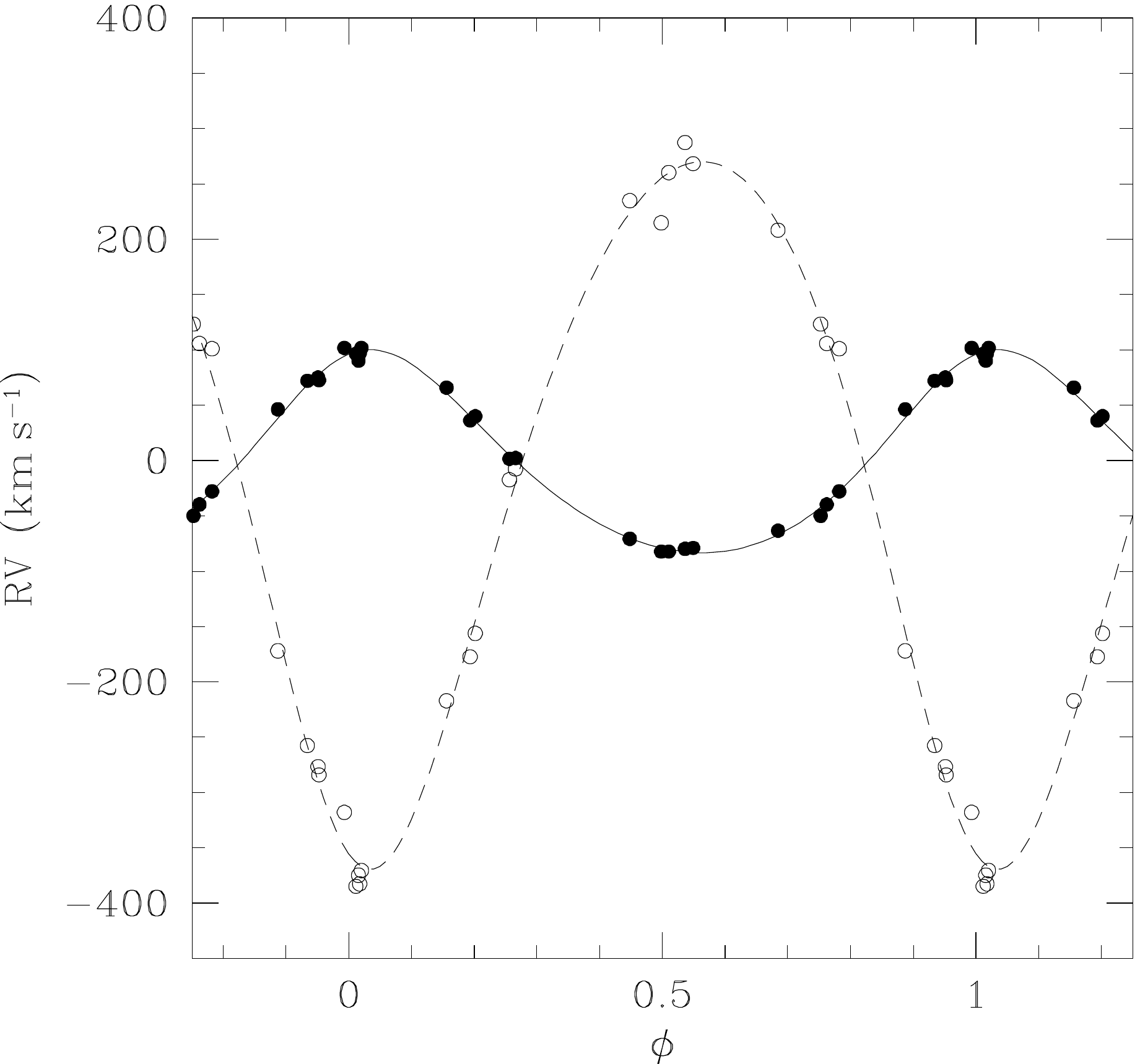}}
\end{minipage}
\caption{Left: SB1 radial velocity curve of Tr\,16-112 as derived from the RVs of the He\,{\sc ii} lines in the primary spectrum. The filled circles stand for the FEROS data, whilst the open triangles indicate RVs derived from the Casleo echelle spectra. Right: SB2 radial velocity curve of Tr\,16-112 as derived from the RVs obtained through the disentangling of the FEROS spectra. Filled symbols stand for the primary whilst the open symbols indicate the secondary star. The orbital solutions shown are described in Table\,\ref{solorb}. Phase $\phi = 0.0$ corresponds to the periastron passage.\label{RVcurve}}
\end{figure*}
As a next step, we used the Li\`ege Orbital Solution Package (LOSP) based on the method of \citet*{WHS} revised by \citet*{Sana06}, to compute the SB1 and SB2 orbital solutions of Tr\,16-112. For the SB1 solution, we used the RVs of the primary's He\,{\sc ii} lines as measured on the entire set of high-resolution echelle spectra, whilst we adopted the RVs derived from the disentangling method applied to the FEROS spectra for the SB2 solution. The corresponding RV curves are shown in Fig.\,\ref{RVcurve} and the parameters of the orbital solutions are listed in Table\,\ref{solorb}. Note that leaving the orbital period as a free parameter did not change the quality of the orbital solution.

The mass ratio (primary/secondary) is found to be $3.49 \pm 0.08$. To our knowledge, this is one of the largest mass ratios found in an OB binary system to date. Our orbital solution confirms the existence of a significant eccentricity, although it is less extreme than previously thought. Note that most parameters agree reasonably well between the SB1 and SB2 orbital solutions, except for the longitude of periastron which has a rather large uncertainty. Comparing the relatively large minimum mass of the secondary with the typical masses of B2 stars suggests that the orbital inclination could be as large as $60^{\circ}$. Under these circumstances, grazing eclipses could {\it a priori} be possible. 

\subsection{Photometric data}
Our photometric data reveal that Tr\,16-112 displays low level variations in its light curve with an amplitude of about 0.03\,mag. When folded with the ephemeris found from the radial velocity curve, we note that our observations actually miss the phase when secondary eclipse would be expected (around $\phi = 0.842$) but cover the phase where primary eclipse would occur ($\phi = 0.251$). The photometric data reveal no trace of an eclipse at this orbital phase and we thus conclude that Tr\,16-112 does not display photometric eclipses. Rather the low-level photometric variability must be due to the ellipsoidal shapes of the stars. 

We analysed the light curve of this system by means of the {\sc nightfall} code\footnote{ http://www.hs.uni-hamburg.de/DE/Ins/per/Wichmann/\\Nightfall.html} developed and maintained by R.\ Wichmann, M.\ Kuster and P.\ Risse. This software uses a Wilson-Devinney like iterative approach to solve the light curves of close binary systems in the Roche geometry, generalized to deal with the more complex situation of eccentric binaries where the sizes of the Roche lobes (and hence the values of the filling factors) change with orbital phase (see \citealt{Kopal}). Despite the limited phase coverage and the small amplitude of the variations, we can attempt to get a decent fit of the data by varying the relevant model parameters which are in this case the orbital inclination, as well as the primary and secondary filling factors, defined as the ratio between the polar radius of the star and the polar radius of the Roche lobe at periastron passage.

We set the effective temperature of the primary star to 39000\,K \citep*{Martins} whilst the secondary temperature was fixed to 22000\,K according to the \citet{SK} calibration, which seems reasonable for B2 main-sequence stars \citep{FM05}. The mass ratio, the orbital eccentricity as well as the longitude of periastron were fixed at the values derived from the radial velocity curve (see Table\,\ref{solorb}). The {\sc nightfall} code was applied to the data in the $V$-band since they are more numerous (996 data points) and of the highest quality. Due to the limited phase coverage of the data and the small range of variability, the parameters are actually not that well constrained. To evaluate the uncertainties on the model parameters, we systematically explored the parameter space, fixing one parameter at once, and searching for the combinations of the other two parameters that yield the lowest residuals.  

The best fit value of the orbital inclination is found to be $54^{\circ}$. This then corresponds to absolute masses of ($41.0 \pm 2.5$)\,M$_{\odot}$ and ($11.7 \pm 0.6$)\,M$_{\odot}$ for the primary and secondary respectively. 

Given the rather large difference in temperature (and thus in surface brightness) between the primary and the secondary, the fit of the light curve is not very sensitive to the value of the secondary filling factor. As a result, the light curve does not allow us to effectively constrain the brightness ratio of the two stars independently of the results obtained from the spectroscopy. The situation is better for the primary (which dominates the light of the system). For this component, we can constrain the filling factor to be in the range 0.72 to 0.74 with a preferred value of 0.73.

\begin{table}
\caption{Parameters of the best fit model for the light curve of Tr\,16-112. The error bars correspond to the 90\% confidence intervals. The average residual of the fit amounts to $\overline{|O - C|} = 0.005$\,mag. \label{lc}}
\begin{tabular}{l c c}
\hline
                       & Primary        & Secondary \\
\cline{2-3}
$i$ ($^{\circ}$)       & \multicolumn{2}{c}{$54^{+4}_{-3}$}\\
$T_{\rm eff}$ (K)      & $39000$ (fixed)& $22000$ (fixed) \\
Periastron Roche lobe filling factor & $0.73^{+0.01}_{-0.02}$ & $0.55^{+0.09}_{-0.11}$ \\
$m$ (M$_{\odot}$)      & $41.0 \pm 2.5$ & $11.7 \pm 0.6$ \\
mean radius $R$ (R$_{\odot}$) & $13.5 \pm 0.6$ & $5.7 \pm 1.1$ \\
$\log{L/L_{\odot}}$    & $5.57 \pm 0.04$ & $3.83 \pm 0.16$ \\
\hline
\end{tabular}
\end{table}  

\begin{figure}
\resizebox{8.5cm}{!}{\includegraphics{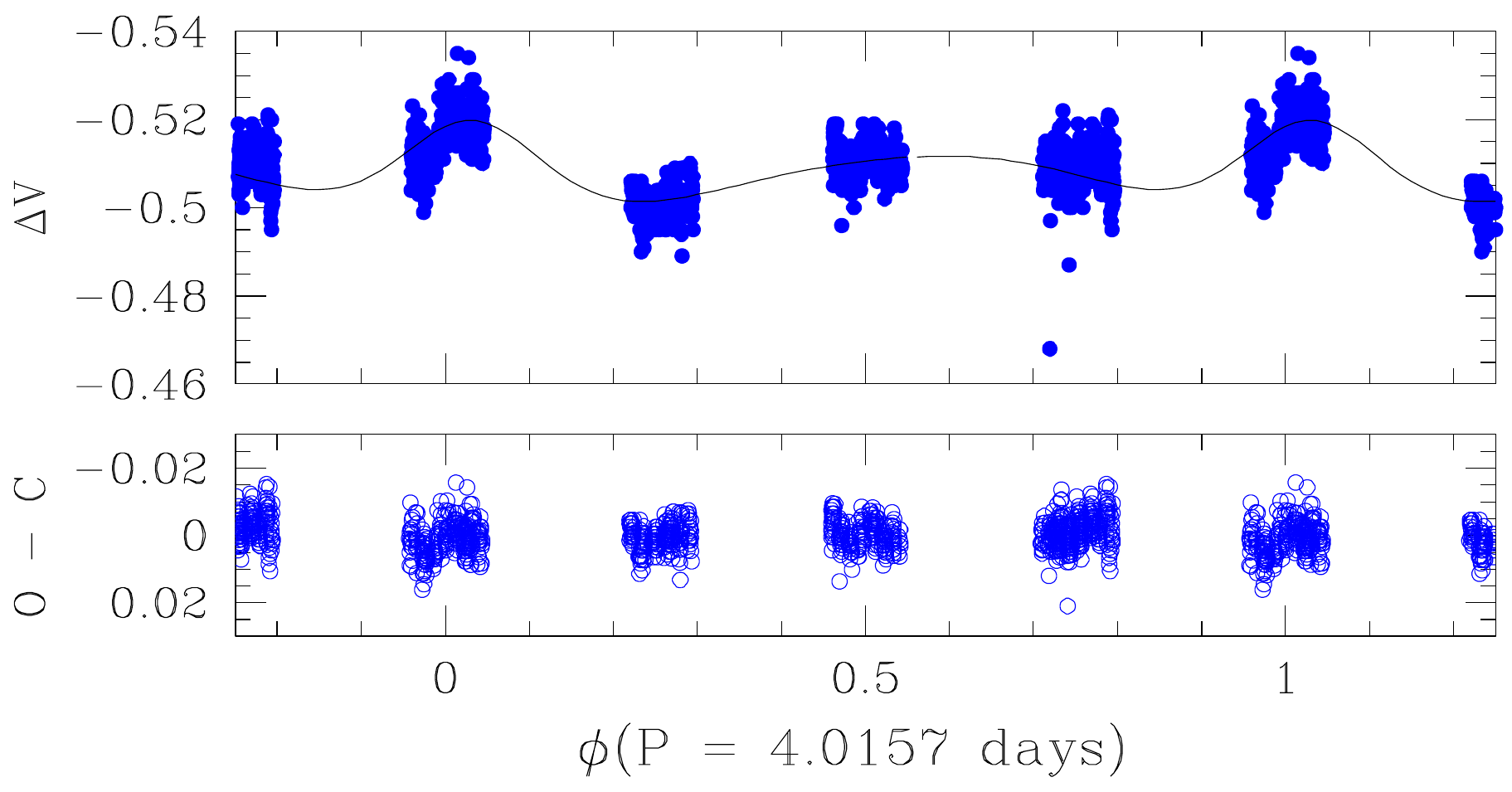}}
\caption{Light curve of Tr\,16-112 in the $V$-band (top panel). The solid curve yields the best fit model where the variations are due to the ellipsoidal shape of the primary star filling part of its critical volume in an eccentric binary. The parameters of the fit are listed in Table\,\ref{lc}. The residuals are shown in the bottom panel.\label{LC}}
\end{figure}

We emphasize that the parameters in Table\,\ref{lc} should be taken as a first estimate only. A more extensive, higher quality (ideally space-borne), photometric campaign would be needed to fill in the gaps in the light curve and ascertain these numbers. Here, we only wish to stress that the observed light curve can be explained through ellipsoidal variations. Pending confirmation of these parameters by future studies, we note that the primary is slightly more massive, significantly larger and brighter than a `typical' main sequence star of the same spectral type. In fact, these parameters would be more representative of luminosity class III objects \citep{Martins}, although this would be at odds with the presence of the strong He\,{\sc ii} $\lambda$\,4686 absorption in the primary spectrum. This would then suggest that Tr\,16-112 is actually a slightly evolved object, contrary to systems such as Tr\,16-104 \citep{GR01} or FO\,15 \citep{Niemela}. Another point worth mentioning is that if the above parameters are indeed confirmed and if the reddening law towards this star is characterized by the typical value of the Trumpler\,16 cluster ($R_V = 3.48 \pm 0.33$, see \citealt{CAR04}), then the distance modulus of Tr\,16-112 would be $13.0 \pm 0.3$ (i.e.\ corresponding to a distance of 4.0\,kpc), almost twice the distance inferred from the study of several eclipsing binaries in Tr\,16 (e.g.\ Tr\,16-1, \citealt{Freyhammer} and Tr\,16-104, \citealt{GR01}). Whilst this is a surprising result, we have to recall that our line of sight towards the Carina region is actually roughly along the Sagittarius-Carina spiral arm, so that several distinct populations of massive stars could be seen projected on the same area of the sky. In the case of the eclipsing binary FO\,15 \citep{Niemela}, a similar contradiction was solved by showing that the latter star is subject to a highly anomalous extinction with $R_V = 4.15$. We note that for Tr\,16-112, there is currently no evidence for such a large value of $R_V$. In addition, since the $E(B - V)$ colour excess of Tr\,16-112 is much lower than for FO\,15, adopting $R_V = 4.15$ for Tr\,16-112 results in a reduction of the distance modulus of 20\% only.

\section{HD\,93343, a peculiar binary system}
Based on 10 low-dispersion spectra, \citet{Levato} reported RV variations of HD\,93343 (O8\,Vn) although they could not provide firm evidence for the multiplicity of the star. Later on, an abstract reporting a very preliminary SB1 orbital solution ($P_{\rm orb} \sim 44$\,days, $e \sim 0.39$, $K_1 \simeq 71$\,km\,s$^{-1}$) was presented by \citet{SN}. 

\begin{figure}
\resizebox{8cm}{!}{\includegraphics{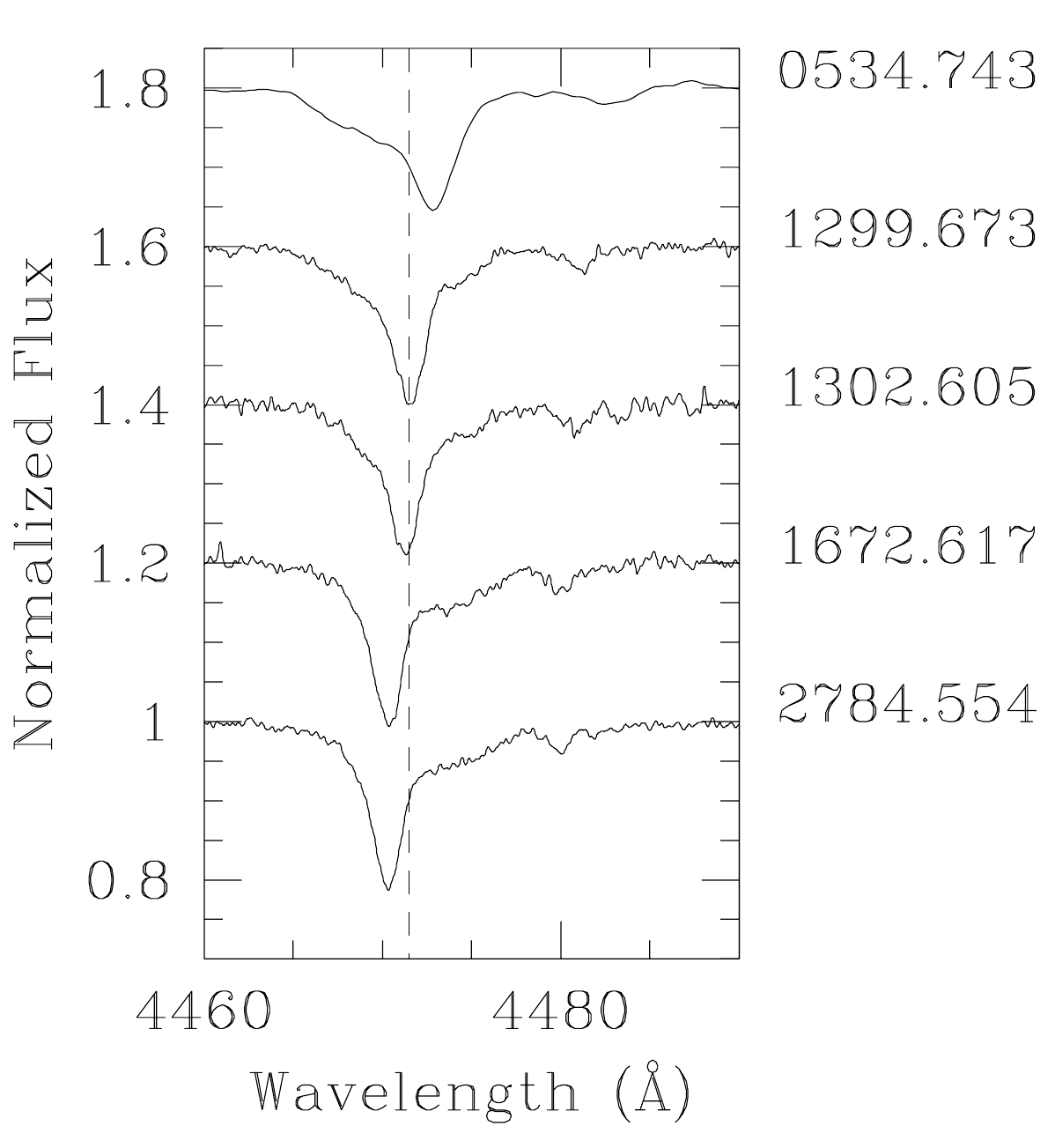}}
\caption{Montage of several of the He\,{\sc i} $\lambda$\,4471 profiles observed in the spectrum of HD\,93343. The date of the observation is indicated on the right in the format HJD$-$2450000.\label{montage}}
\end{figure}

Here, we analyse six medium-resolution Boller \& Chivens and nine high-resolution FEROS spectra of HD\,93343 spread over six years. These data reveal the SB2 nature of this star: one of the components of HD\,93343 has narrow absorption lines (${\tt FWHM} \sim 125$\,km\,s$^{-1}$), whilst the companion displays much broader lines (${\tt FWHM} \sim 500$\,km\,s$^{-1}$! ). 
As can be seen on Fig.\,\ref{montage}, the sharper lines obviously move in radial velocity. The RV changes of the broad lines are less obvious to see. However simultaneous fits of two Gaussians to the observed profiles clearly indicate that the broad lines move in phase opposition with the narrow lines, although their RVs are affected by large uncertainties (at least 20\,km\,s$^{-1}$) due to their widths. Our RV data allow us to establish a preliminary estimate of the mass-ratio of the system $m_{\rm narrow}/m_{\rm broad} = 0.63 \pm 0.06$. 

The apparent EWs (i.e.\ with respect to the combined primary + secondary continuum) of the He\,{\sc i} $\lambda$\,4471 line as determined from the two Gaussian fits are $0.57 \pm 0.02$ and $0.23 \pm 0.05$\,\AA\ for the broad and narrow line component respectively. For the He\,{\sc ii} $\lambda$\,4542 lines, the apparent EWs are $0.41 \pm 0.05$ and $0.17 \pm 0.04$\,\AA\ respectively. 
From these numbers, we infer a spectral type O8 for both stars ($\log{W'} = 0.14 \pm 0.06$ for the broad line component and $\log{W'} = 0.11 \pm 0.14$ for the star with the sharper lines, see \citealt{Conti73b} and \citealt{Mathys}). We note that the spectral type of the component with the sharper lines could actually be in the range O7-8.5. The same equivalent widths allow us to estimate that the broad line component should be roughly 2.4 times brighter in the optical than the narrow line component. Unfortunately, an attempt to disentangle the spectra of the two stars failed because the RV changes sampled by our data do not cover the full width of the broad lines. In such a situation, the disentangling method cannot recover the spectra of the components (see \citealt{GL06}). 

The likely combination of a slow and fast rotator in HD\,93343 is reminiscent of the situation encountered in Plaskett's Star (HD\,47129, see \citealt{Linder}) and this system could thus provide an interesting test case for binary evolution theories (e.g.\ \citealt{deMink}). Unfortunately, the existing data are not sufficient to establish a full orbital solution. From the RV changes observed over 6 consecutive nights in March 1997 and May 1999 (see Table\,\ref{RV93343}), we can guess that the orbital period is likely of order a few weeks. A very preliminary Fourier analysis indicates the presence of several alias peaks of identical amplitude in the periodogram for periods around 40 -- 50 days, with the highest one corresponding to a period of 44.54\,days. We note that the latter result is in good agreement with the preliminary period reported by \citet{SN}. 

\begin{figure}
\resizebox{8cm}{!}{\includegraphics{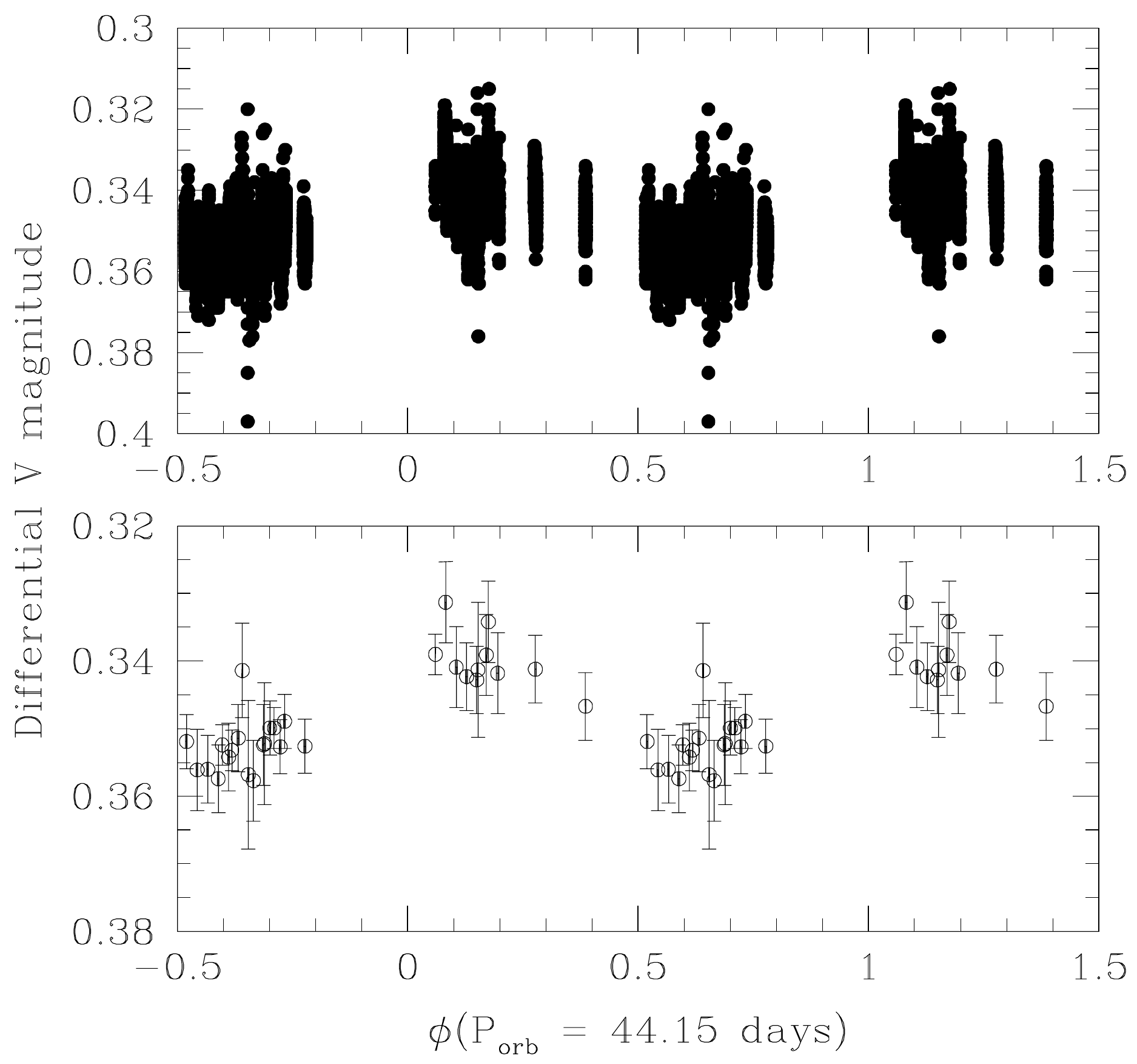}}
\caption{Photometric variations of HD\,93343. The top panel illustrates the raw data ($\Delta\,V$) folded with a period of 44.15\,days. Phase zero was arbitrarily set to HJD\,2450000. The bottom panel yields the same light curve, but where the data of each single observing night have now been averaged. The error bars on the data yield the dispersion of $\Delta\,V$ about the mean for each observing night.\label{lc93343}}
\end{figure}

\begin{table}
\caption{Radial velocities of HD\,93343 as inferred from a simultaneous fit of two Gaussians to the He\,{\sc i} $\lambda\lambda$\,4471, 4921, 5876 and He\,{\sc ii} $\lambda$\,4542 line profiles observed in our data. Note that for the Boller \& Chivens spectra (B\&C), only the He\,{\sc i} $\lambda$\,4471 and He\,{\sc ii} $\lambda$\,4542 lines were measured. \label{RV93343}}
\begin{tabular}{r r r c}
\hline
\multicolumn{1}{c}{HJD$-$2450000} & Narrow lines & Broad lines & Inst.\\
\hline
 534.743 & $-52.5$ & 101.0 & B\&C \\
 535.749 & $-55.8$ &  87.7 & B\&C \\
 536.716 & $-47.9$ &  63.0 & B\&C \\
 537.731 & $-32.2$ &  53.9 & B\&C \\
 538.715 & $-50.9$ &  47.4 & B\&C \\
 539.717 & $-38.3$ &  42.4 & B\&C \\
1299.673 & 4.8 & $-20.6$ & F \\
1300.677 & 0.2 & $-27.4$ & F\\
1301.630 & $-9.5$ & $-19.4$ & F\\
1302.605 & $-22.7$ & $-6.7$ & F\\
1304.650 & $-30.0$ & 13.0 & F\\
1670.632 & $-79.2$ & 70.5 & F\\
1672.617 & $-81.3$ & 60.7 & F\\
2783.550 & $-80.5$ & 49.1 & F\\
2784.554 & $-81.3$ & 59.3 & F\\
\hline
\end{tabular}
\end{table}
We finally analysed an extensive set of 6101 $V$-band photometric measurements of HD\,93343 taken between June 2004 and April 2008 with the same instrumentation as for Tr\,16-112. Although the photometric variations are rather small, the \citet{HMM} periodogram actually presents its highest peak for a period of 44.15\,days with a peak-to-peak amplitude of 0.016\,mag. Whilst another peak with almost identical amplitude is found in the periodogram for a period of 80.3\,days, we nevertheless conclude that the orbital period of HD\,93343 is indeed likely to be around 44\,days. We stress that the light curve folded with a period of 44.15\,days indicates a single modulation per cycle (see Fig.\,\ref{lc93343}) unlike what one would expect for purely ellipsoidal variations assuming the system has a circular orbit. A modelling of these photometric variations must however await the results of a more extensive spectroscopic (and photometric) monitoring campaign of this system.  

\section{The RV constant star HD\,93250}
\begin{figure}
\resizebox{8cm}{!}{\includegraphics{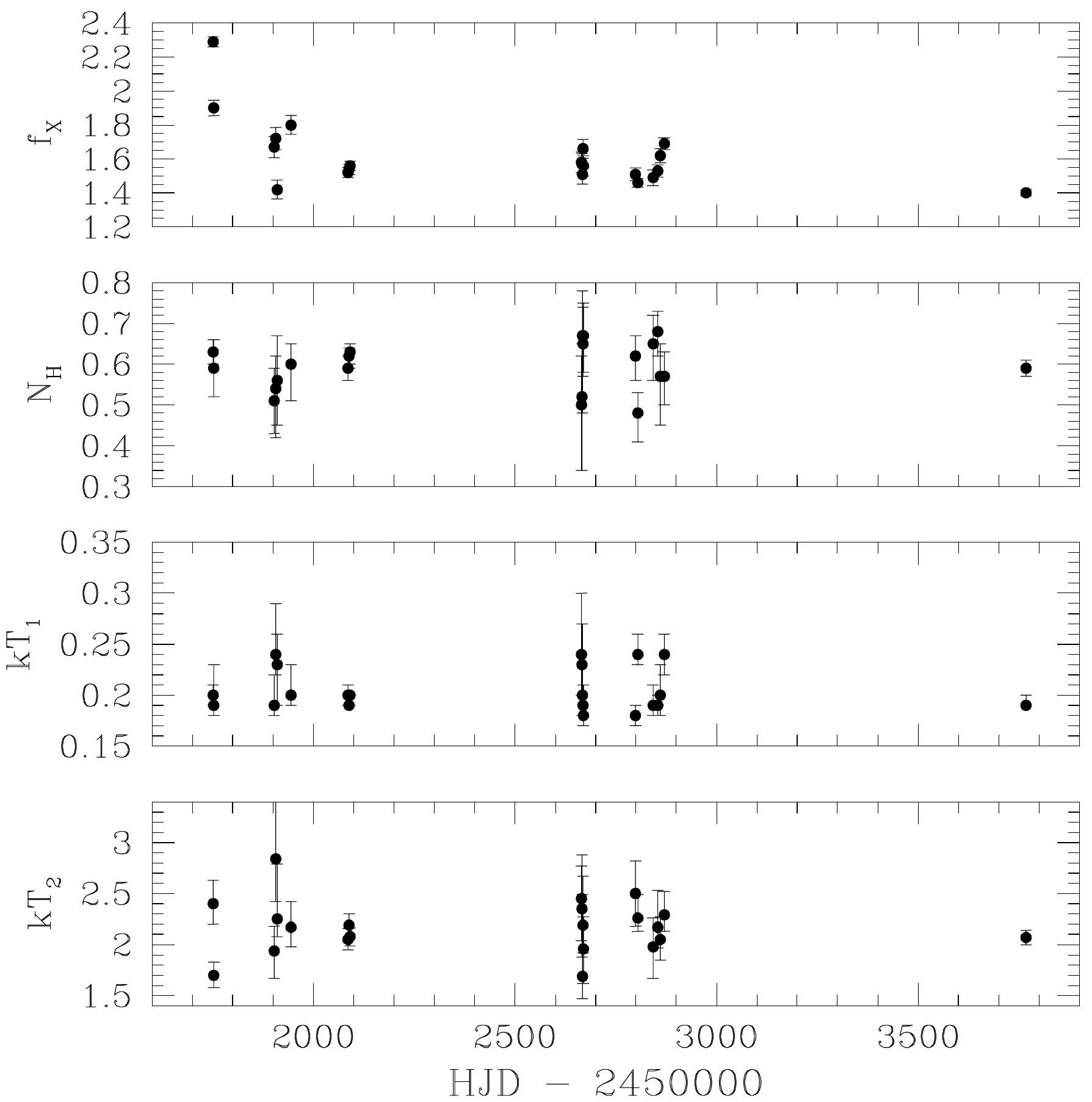}}
\caption{From top to bottom the various panels show the observed X-ray flux of HD\,93250 (in units $10^{-12}$\,erg\,cm$^{-2}$\,s$^{-1}$) over the 0.5 -- 10\,keV domain, the wind hydrogen column density (in units $10^{22}$\,cm$^{-2}$) in addition to the interstellar column of $2.55 \times 10^{21}$\,cm$^{-2}$, the temperatures of the first and second thermal component (in keV). \label{lcX}}
\end{figure}

Last, but not least, HD\,93250 (O3.5, \citealt{O2}) was considered to have a constant radial velocity by \citet{Levato}. This statement was based on 9 low-resolution spectra obtained on 9 consecutive nights. The 1-$\sigma$ dispersion of their RV measurements was 11.6\,km\,s$^{-1}$. However, some indirect indications of a possible binarity were found in the radio and X-ray domains. In fact, \citet{LCK} reported on an Australia Telescope Compact Array radio observation of HD\,93250. The star was detected only at the 8.64\,GHz frequency. Assuming the radio flux to be of thermal origin and adopting a distance of 2.2\,kpc, \citet{LCK} inferred a mass loss rate of $\log{\dot{M}} \sim -4.39 \pm 0.15$, considerably larger than the value derived from analyses of the optical spectra of this star\footnote{Indeed, \citet{Martins2} subsequently derived $\log{\dot{M}} = -6.25 \pm 0.7$ from a model atmosphere fit.}. Therefore, \citet{LCK} suggested that the radio emission is essentially non-thermal. Since non-thermal (synchrotron) radio emission is nowadays commonly associated with a wind-wind collision in an early-type binary (see e.g.\ \citealt{michael}), one would thus expect HD\,93250 to be a binary system. Some hints for binarity come also from the X-ray domain. Indeed, {\it Chandra} observations revealed that this star has the largest $\log{L_{\rm X}/L_{\rm bol}}$ ($= -6.5$) among the objects studied in this paper \citep{Sanchawala}. Moreover, the star has been frequently observed with {\it XMM-Newton} in the course of various campaigns on the Carina region. In the framework of a global study of the X-ray properties of OB stars (Naz\'e 2009, in preparation), we have retrieved and analysed 21 {\it XMM-Newton}-EPIC spectra of HD\,93250 from the 2XMM catalogue \citep{Watson}. The X-ray flux of HD\,93250 displays clear variability between the different pointings (see Fig.\,\ref{lcX}), whilst most single O-type stars usually exhibit a rather constant X-ray flux. The EPIC spectra of HD\,93250 can be fitted with a two-temperature optically thin thermal plasma {\tt mekal} model with temperatures around 0.20\,keV and 1.7 -- 2.8\,keV. These results, i.e.\ the existence of a relatively hard thermal component and the significant X-ray variability could be seen either as further evidence for HD\,93250 being a colliding wind binary or alternatively they could indicate that this star harbours a magnetically confined wind similar to $\theta^1$\,Ori\,C \citep{Gagne}.\\  

Six FEROS high-resolution spectra of HD\,93250 were analysed. The spectrum of this star displays several emission lines: N\,{\sc iii} $\lambda\lambda$\,4634-40, C\,{\sc iii} $\lambda\lambda$\,5696, 6721, 6727-31 and Si\,{\sc iv} $\lambda\lambda$\,4089, 4116 are the most prominent ones. A weak N\,{\sc iv} $\lambda$\,4058 emission is also seen, whilst He\,{\sc ii} $\lambda$\,4686 is in strong absorption ($EW = 0.54 \pm 0.06$\,\AA) with a slightly asymmetric profile. These features indicate that HD\,93250 is an O\,V((f$^+$)) star. \citet{O2} introduced the new O3.5 spectral type and proposed HD\,93250 to be a prototype of this category. This result is in good agreement with the ratio of the equivalent widths of the He\,{\sc i} $\lambda$\,4471 ($EW = 0.119 \pm 0.008$\,\AA) and He\,{\sc ii} $\lambda$\,4542 ($EW = 0.690 \pm 0.018$\,\AA) classification lines. The \citet{Conti73b} criterion yields $\log{W'} = -0.76 \pm 0.03$ corresponding to an O4 spectral type\footnote{Note that \citet{Conti73b} did not consider an O3.5 spectral type in his classification scheme.}. In summary, HD\,93250 hence appears to be an O3.5\,V((f$^+$)) star.
 
Our RV measurements indicate 1-$\sigma$ RV dispersions of less than 2.0\,km\,s$^{-1}$ for the He\,{\sc i} $\lambda$\,4471, He\,{\sc ii} $\lambda\lambda$\,4542 and 5412 absorption lines. The same conclusion holds if we consider the mean RV of seven lines ($\sigma = 1.3$\,km\,s$^{-1}$, see Table\,\ref{RV93250}). Note that the largest variability ($\sigma = 6.5$\,km\,s$^{-1}$) is found for the He\,{\sc ii} $\lambda$\,4686 line. This is no surprise since this line shows some hints for a P-Cygni type profile and the variability is thus most likely due to variations of the stellar wind. For comparison, the RV dispersion of the diffuse interstellar band at 5780\,\AA\ amounts to 0.7\,km\,s$^{-1}$. Therefore, we conclude that no significant RV variations are observed for the photospheric lines of  HD\,93250. Our measurements sample time scales of 3 - 5 days as well as of several years. The lack of significant RV variations therefore indicates that HD\,93250 is most likely not a spectroscopic binary with an orbital period of order a few days to a few weeks or a few years. Finally, it should be noted that \citet{Mason} did not report a visual companion to HD\,93250. In conclusion, HD\,93250 is probably either a single star or a moderately wide binary system seen under a low inclination angle. 
\begin{table}
\caption{Radial velocities of HD\,93250 as measured from the He\,{\sc i} $\lambda\lambda$\,4471, 5876, He\,{\sc ii} $\lambda\lambda$\,4026, 4200, 4542, 5412 and O\,{\sc iii} $\lambda$\,5592 absorption lines.\label{RV93250}}
\begin{tabular}{c r r c}
\hline
HJD$-$2450000 & Mean RV\\
\hline
1304.551 &  $-9.9$ \\
2782.538 &  $-9.1$ \\
2784.526 & $-12.5$ \\
3130.501 &  $-9.2$ \\
3131.455 &  $-9.3$ \\
3133.454 & $-10.5$ \\
\hline
\end{tabular}
\end{table}

\section{O-type binaries in and around the Trumpler\,16 cluster}
In this section, we briefly discuss some of the results obtained so far in the course of the optical campaign of the {\it X-Mega} project. In addition to the stars studied in this paper, we include the results on some other stars: Tr\,16-34, HD\,93205, Tr\,16-104, Tr\,16-110, HD\,93161, FO\,15 taken from the previous papers of this series \citep{ALB01,ALB02,Morrell,GR01,Naze,Niemela} as well as Tr\,16-1 and HD\,93403 from the work by \citet{Freyhammer} and \citet{GR00} respectively.

The properties of the eclipsing systems, Tr\,16-1, Tr\,16-104 and FO\,15 \citep{Freyhammer,GR01,Niemela}, are particularly useful since their light curve and orbital solution provide absolute parameters and allow us to constrain their distance. Indeed, the investigations of these three eclipsing systems agree with a distance of about 2.5\,kpc. For the ellipsoidal variables, Tr\,16-112 and HD\,93205, the uncertainties on the orbital inclination and hence on the absolute parameters are much larger (\citealt{Antokhina} and this work).

For the non-eclipsing binaries and for the probably single star HD\,93250, we have assumed a distance of 2.5\,kpc. For each star, we have evaluated the effective temperature and the bolometric corrections according to \citet{Martins} and \citet{MP}. We have generally assumed an uncertainty of half a spectral type. To derive the luminosities, we need the observed magnitude, the reddening, the brightness ratios between the different components, the bolometric corrections and an estimate of the distance. The observed magnitudes and colours were taken from the WEBDA database. The relative brightness of the components of the binary systems was estimated in the various papers from the ratio of the EWs of the primary and secondary lines. The reddening was evaluated according to the results of \citet{CAR04}. These authors argued that a unique reddening law is not appropriate to study the whole Carina region and they estimated different values of the selective extinction $R_V = A_V/E(B - V)$ for each cluster. 

This approach yields the bolometric luminosities quoted in Table\,\ref{tr16hrd}. For Tr\,16-112, we use the luminosities derived from the light curve analysis, rather than assuming a distance of 2.5\,kpc. The luminosity of HD\,93161B might be overestimated if the star turned out to be indeed another binary system (see \citealt{Naze}). Note that for HD\,93403, the evolutionary masses are revised downwards compared to the previous analysis \citep{GR00} and the evolutionary mass ratio is now in better agreement with the observed value ($1.75 \pm 0.06$). Finally, we stress that the bolometric luminosity of HD\,93250 that we obtain through our approach ($\log{\rm L/L_{\odot}} = 6.09 \pm 0.07$) is in very good agreement with the luminosity $\log{\rm L/L_{\odot}} = 6.12^{+.25}_{-.17}$ inferred by \citet{Martins2} using a different method. 

\begin{figure}
\begin{center}
\resizebox{8cm}{8cm}{\includegraphics{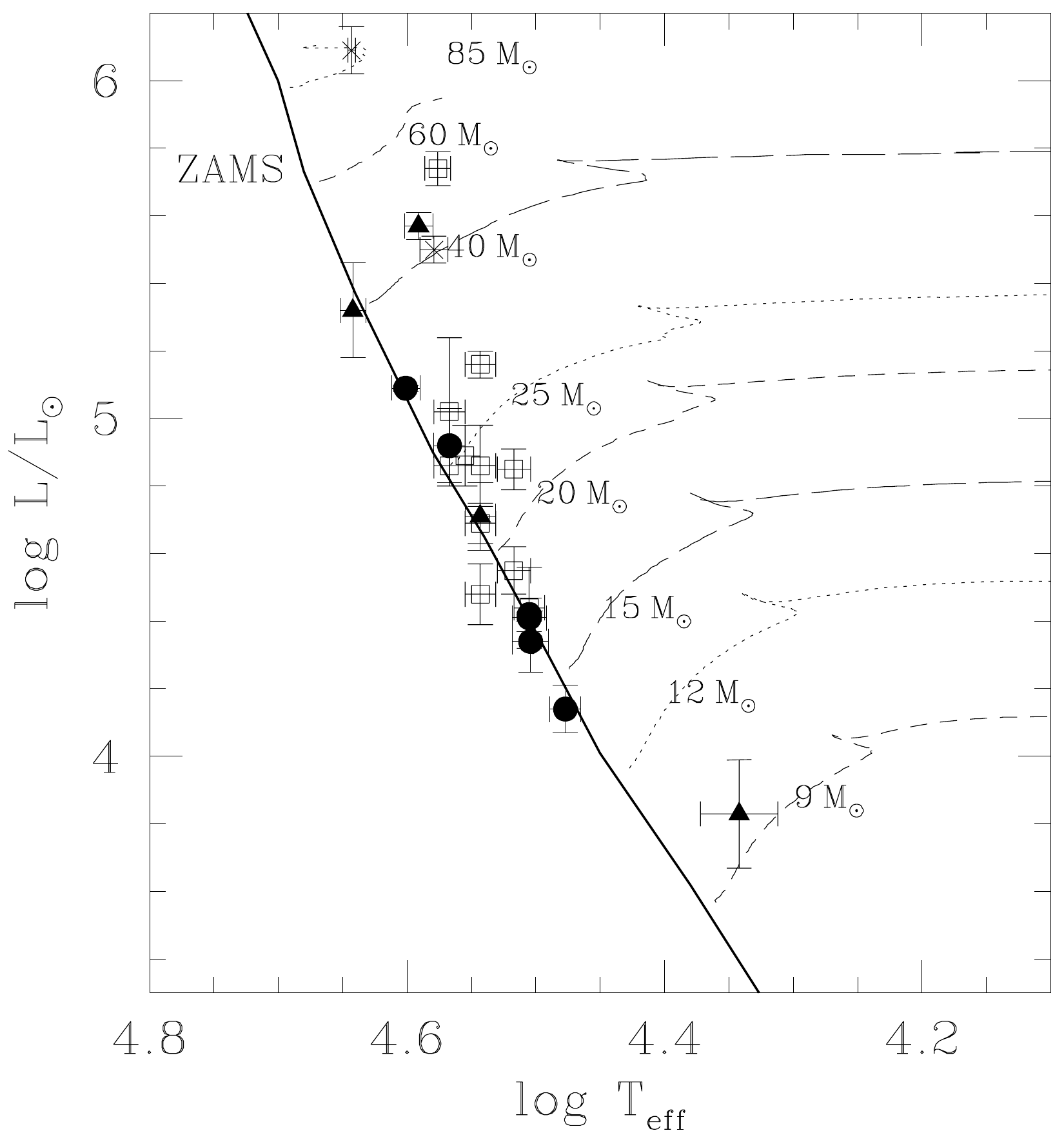}}
\caption{Location in the Hertzsprung-Russell diagram of the components of various early-type systems of the Carina region. Eclipsing binaries are shown by filled dots, whilst ellipsoidal variables are shown as filled triangles. Non-eclipsing binary systems are indicated by open squares and the asterisks stand for two probably single stars. The single star evolutionary tracks from \citet{MM} for an initial rotational velocity of 300\,km\,s$^{-1}$ at solar metallicity are shown for those parts of the tracks where the surface hydrogen abundance $X \geq 0.4$.
\label{hrdtr16}}
\end{center}
\end{figure}

In Fig.\,\ref{hrdtr16}, we have plotted the various stars of Table\,\ref{tr16hrd} in a Hertzsprung-Russell diagram along with the evolutionary tracks from \citet{MM} for solar metallicity. Provided that all stars (except maybe Tr\,16-112) are indeed at the same distance from Earth, this diagram suggests that the components of HD\,93161 and HD\,93403 are evolved off the zero age main-sequence with an age of about 2 -- 3\,Myr, unlike the components of the eclipsing binaries and of Tr\,16-110 which are found very close to the zero-age main-sequence (ZAMS). 

\begin{table*}
\caption{Orbital properties of the short-period spectroscopic binaries in the Carina complex. For the third component of the triple system Tr\,16-104, the two possible solutions are listed (see \citealt{GR01}). In the last column, SB1 and SB2 stand respectively for a single-lined and a double-lined spectroscopic binary, whilst the e and v letters indicate systems that are known to display photometric eclipses or ellipsoidal variations respectively.\label{tr16}}
\small
\begin{center}
\begin{tabular}{l c c c c c c c}
\hline
System & spectral types & P$_{\rm orb}$ & $e$ & $\omega$ & $q = m_2/m_1$ & Type \\
& & (days) & & ($^{\circ}$) & \\
\hline\hline
\vspace*{-3mm} & & & & & & & \\
Tr\,16-1   & O9.5\,V + B0.3\,V & 1.4693 & 0.0              & --             & $0.83 \pm 0.08$ & SB2e \\
Tr\,16-34  & O8\,V + O9.5\,V   & 2.3000 & 0.0              & --             & $0.76 \pm 0.01$ & SB2  \\
HD\,93205  & O3\,V + O8\,V     & 6.0803 & $0.371 \pm 0.005$& $50.6 \pm 0.9$ & $0.42 \pm 0.01$ & SB2v \\
\vspace*{-3mm} & & & & & & \\ 
\hspace*{-3mm} $\begin{array}{l} {\rm Tr\,16-104}\end{array} $ & $\begin{array}{l} {\rm O7\,V + O9.5\,V + B0.2\,IV}\end{array} $ & $\left\{ \begin{array}{c} {2.1529} \\ {285 , 1340} \end{array} \right.$ & $\begin{array}{c} {0.0} \\ {0.25 \pm 0.05 , 0.37 \pm 0.04} \end{array}$ & $\begin{array}{c} {-} \\ {287 \pm 20 , 238 \pm 8} \end{array}$ 
& $ \begin{array}{l} {0.65 \pm 0.03} \\ {} \end{array}$ & $\begin{array}{c} {\rm SB2e} \\ {\rm SB1} \end{array}$ \\
\vspace*{-3mm} & & & & & & \\ 
\hspace*{-3mm} $\begin{array}{l} {\rm Tr\,16-110}\end{array} $ & $\left\{ \begin{array}{c} {\rm O7\,V + O8\,V} \\ {\rm O9\,V} \end{array} \right.$ & $\begin{array}{c} {3.6284} \\ {5.034} \end{array}$ & $\begin{array}{c} {0.06 \pm 0.03} \\ {0.09 \pm 0.04} \end{array}$ & $\begin{array}{c} {355 \pm 15} \\ {209 \pm 30} \end{array}$ & $ \begin{array}{l} {0.96 \pm 0.04} \\ {} \end{array}$ & $\begin{array}{c} {\rm SB2} \\ {\rm SB1} \end{array}$ \\
\vspace*{-3mm} & & & & & & \\ 
HD\,93161A & O8\,V + O9\,V     & 8.566  & 0.0              & --             & $0.76 \pm 0.01$ & SB2 \\
FO\,15     & O5.5\,V + O9.5\,V & 1.1414 & 0.0              & --             & $0.52 \pm 0.04$ & SB2e  \\
Tr\,16-112 & O5.5\,V + B2\,V-III & 4.0157 & $0.15 \pm 0.01$  & $342.8 \pm 5.6$& $0.29 \pm 0.01$ & SB2v  \\
HD\,93403  & O5.5\,I + O7\,V   & 15.093 & $0.23 \pm 0.02$  & $22.5 \pm 4.4$ & $0.57 \pm 0.02$ & SB2  \\
\hline
\end{tabular}
\end{center}
\end{table*}

\begin{table}
\caption{Revised temperatures and bolometric luminosities of the components of eclipsing binaries (top part), the ellipsoidal variables (second part), the non-eclipsing binaries (third part) and two presumably single O-stars (lower part) in the Carina complex.\label{tr16hrd}}
\begin{center}
\begin{tabular}{l c c c}
\hline 
\multicolumn{1}{c}{System} & Spectral type & T$_{\rm eff}$ (K) & $\log{\rm L/L_{\odot}}$ \\
\hline\hline
Tr\,16-1a   & O9.5\,V    & $32000 \pm 500$  & $4.42 \pm 0.06$ \\
Tr\,16-1b   & B0.3\,V    & $30000 \pm 800$  & $4.14 \pm 0.07$ \\
Tr\,16-104a & O7\,V      & $36900 \pm 1000$ & $4.92 \pm 0.11$ \\ 
Tr\,16-104c & O9.5\,V    & $31900 \pm 1000$ & $4.34 \pm 0.09$ \\
FO\,15a     & O5.5\,V    & $40000 \pm 1000$ & $5.09 \pm $ 0.01\\
FO\,15b     & O9.5\,V    & $32000 \pm 500$  & $4.41 \pm $ 0.01\\
\hline
HD\,93205a  & O3.5\,V    & $43900 \pm 1000$ & $5.32 \pm 0.14$ \\
HD\,93205b  & O8\,V      & $34900 \pm 1000$ & $4.71 \pm 0.10$ \\
Tr\,16-112a & O5.5-6\,V  & $39000 \pm 1000$ & $5.57 \pm $ 0.04\\
Tr\,16-112b & B2\,V-III  & $22000 \pm 1500$ & $3.83 \pm $ 0.16\\
\hline
HD\,93161Aa & O8\,V      & $34900 \pm 1000$ & $5.16 \pm 0.04$ \\
HD\,93161Ab & O9\,V      & $32900 \pm 1000$ & $4.85 \pm 0.06$ \\
HD\,93343a  & O8\,V      & $34900 \pm 1000$ & $4.86 \pm 0.12$ \\ 
HD\,93343b  & O7-8.5\,V  & $34900 \pm 1000$ & $4.48 \pm 0.12$ \\
HD\,93403a  & O5.5\,I    & $37700 \pm 900$  & $5.72 \pm 0.05$ \\
HD\,93403b  & O7\,V      & $36900 \pm 1000$ & $4.97 \pm 0.22$ \\
Tr\,16-34a  & O7.5\,V    & $35900 \pm 1000$ & $4.89 \pm 0.09$ \\
Tr\,16-34b  & O9\,V      & $32900 \pm 1000$ & $4.44 \pm 0.12$ \\
Tr\,16-110a & O7\,V      & $36900 \pm 1000$ & $4.86 \pm 0.06$ \\
Tr\,16-110b & O8\,V      & $34900 \pm 1000$ & $4.69 \pm 0.06$ \\
Tr\,16-110c & O9\,V      & $32900 \pm 1000$ & $4.55 \pm 0.07$ \\
\hline
HD\,93161B  & O6.5\,V(f)       & $37900 \pm 1000$ & $5.50 \pm 0.04$ \\
HD\,93250   & O3.5\,V((f$^+$)) & $44000 \pm 300$  & $6.09 \pm 0.07$ \\
\hline
\end{tabular}
\end{center}
\end{table}

\begin{figure}
\begin{center}
\resizebox{8cm}{8cm}{\includegraphics{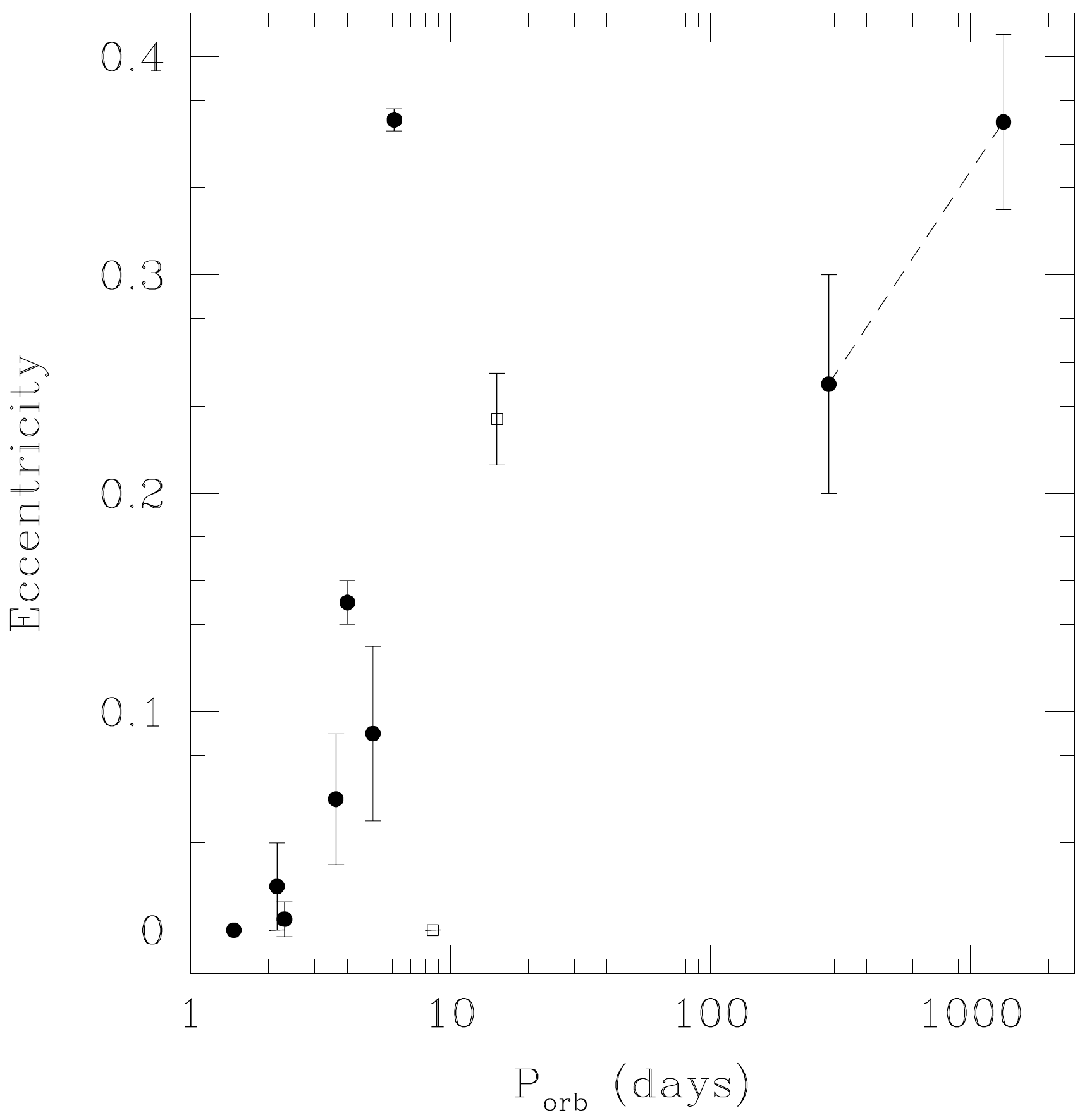}}
\caption{Eccentricity of the orbits of a sample of early-type binaries in the Carina complex as a function of the orbital period. The probable members of Tr\,16 are shown by filled symbols, whereas the two binary systems outside the cluster (HD\,93403 and HD\,93161A) are indicated by open symbols. For the third component of the triple system Tr\,16-104, the two possible solutions are shown, connected by a dotted line. \label{ePorb}}
\end{center}
\end{figure}

Using all the available orbital solutions for early-type binary systems in or around Trumpler\,16, we constructed the period --  eccentricity diagram shown in Fig.\,\ref{ePorb}. This figure confirms our assertion from Fig.\,\ref{hrdtr16} that HD\,93161A must be older and more evolved than the main-sequence stars in the core of Tr\,16. In fact, for the latter, significant eccentricities are found for orbital periods as short as $\sim 4$\,days, indicating that tidal interactions have not yet had the time to circularize the eccentric orbits.

At least three systems of multiplicity three or higher were found among the early-type binaries in the Carina complex. Concerning the fraction of systems of higher multiplicity, \citet{Tokovinin} found that about 43\% of the nearby low-mass (0.5 - 1.5\,M$_{\odot}$) spectroscopic binaries with $P < 10$\,days quoted in the Eighth Catalogue of Spectroscopic Binaries have indeed a tertiary component. The third component provides a natural sink for the angular momentum that needs to be removed from a binary system in order to make it a close one. \citet{Tokovinin} quotes an empirical stability limit for a triple system:
$$P_{\rm out}\,(1 - e_{\rm out})^3 > 5\,P_{\rm in}$$ 
where $P_{\rm in}$, $P_{\rm out}$ and $e_{\rm out}$ correspond respectively to the orbital period of the close binary, the orbital period of the third star around the centre of mass of the close binary and the eccentricity of the third star's orbit. Both possible orbital solutions for the third component in Tr\,16-104 largely satisfy this criterion, whilst the orbital period of the SB1 component in Tr\,16-110 assumed to correspond to $P_{\rm out}$ of a triple system obviously cannot meet this condition, thus confirming the idea that there should exist a fourth (unseen) component in this system.    

\section{Conclusions}
In this paper, we have presented the very first SB2 orbital solution of the Tr\,16-112 binary system. We derived a rather large mass ratio and confirmed that this binary has an orbit with a significant eccentricity despite of its short orbital period. This system displays ellipsoidal variations that can be explained by the primary filling about 73\% of its Roche lobe at periastron. Furthermore, we have presented the first clear evidence for an SB2 signature in the spectrum of HD\,93343. This system likely harbours one fast and one slow rotator and could currently be a post-Roche lobe overflow evolutionary stage, making it a somewhat less extreme version of Plaskett's Star. Finally, we have shown that the properties of the components of most of the binaries in the Carina region (in or around Trumpler\,16) suggest that these are very young objects where tidal interactions have not yet had the time to circularize the orbits. One point that remains to be studied in the forthcoming years is the lack of a clear binary signature in HD\,93250. This system displays a non-thermal radio emission and such a phenomenon is a priori not expected for a single star. Interferometric observations could help to search for a companion that might be either too faint or too far away from the O3.5 primary to be detected in spectroscopy. 

\section*{Acknowledgments}
We thank the referee of this paper for his constructive comments. The Li\`ege team acknowledges multiple support from the Fonds de Recherche Scientifique (FRS/FNRS), through the XMM/INTEGRAL PRODEX contract as well as by the Communaut\'e Fran\c caise de Belgique - Action de recherche concert\'ee - Acad\'emie Wallonie - Europe.

\bsp

\label{lastpage}

\end{document}